\documentclass[a4paper,11pt]{article}

\usepackage{fullpage}
\usepackage[T1]{fontenc}

\usepackage{amssymb,amsmath}

\usepackage{graphicx}

\usepackage{amsmath,amsfonts,epsfig,mathrsfs,todonotes,yfonts,cite}  
\newcommand{\nn}{\nonumber}

\newcommand{\re}[1] {(\ref{#1})}

\newcommand{\bea}{\begin{eqnarray}}
\newcommand{\eea}{\end{eqnarray}}

\newcommand{\al}{\alpha}
\newcommand{\be}{\beta}

\begin{document}

\setcounter{footnote}{0}

\baselineskip 6 mm

\begin{titlepage}
	\thispagestyle{empty}
	\begin{flushright}
		
	\end{flushright}
\rightline{{UUITP-28/25}}

	\vspace{35pt}
	
	\begin{center}
	    { \Large{ \bf Super Yang--Mills with partially non-linear \\[0.3cm] extended supersymmetry}}  
        %Super Yang Mills with partially broken extended supersymmetry
        %Non-abelian theories with partial non-linear supersymmetry 
		
		\vspace{50pt}
		
		{Fotis Farakos$^{a}$ and Ulf Lindstr\"om$^{b,c}$}   
		
		\vspace{25pt}
		
        $^{(a)}${\it Physics Division, National Technical University of Athens \\
        15780 Zografou Campus, Athens, Greece}
		
		\vspace{15pt}

        $^{(b)}${\it Department of Physics and Astronomy, Division of Theoretical Physics, Uppsala University, \\ 
        Box 516, SE-75120 Uppsala, Sweden}
		
		\vspace{15pt}

        $^{(c)}${\it Center for Geometry and Physics, Uppsala University, Box 480, SE-75106 Uppsala, Sweden}
		
		\vspace{15pt}        
		
		\vspace{40pt}
		
		{ABSTRACT} 
	\end{center}

We discuss 4D N=2 non-abelian gauge theories where one supersymmetry is preserved while the other one is spontaneously broken and non-linearly realized. The goldstino resides in a Maxwell multiplet of the Bagger--Galperin type. We introduce appropriate constraints that eliminate the chiral N=1 superfield sector of the non-abelian N=2 multiplets and discuss the properties of the leading order Lagrangians.

\vspace{10pt}

\bigskip

\end{titlepage}

\baselineskip 6mm

\newpage

%\tableofcontents

\section{Introduction}

A striking property of supersymmetry is that it is inherent in string theory both in the linear and in the non-linear version (see e.g. \cite{Ibanez:2012zz,Mourad:2017rrl}). 
The two versions can also co-exist in extended objects that preserve half of the supersymmetry and spontaneously break the other half \cite{Hughes:1986dn,Hughes:1986fa}; 
Objects with such properties are the D-branes, 
which play a pivotal role in our current understanding of string theory \cite{Polchinski:1995mt}. 
Understanding of the properties of partially non-linearly realized supersymmetry (both global and local) is then central to understanding the dynamics of extended objects in string theory.

Within four-dimensional global supersymmetry, in particular, 
only a few developments in the above mentioned direction have been successful. 
For example it is still an open problem how to build {\it superspace} Lagrangians with more than one linear and one non-linear supersymmetry. 
Some attempts have been made but not without their own shortcomings \cite{Ketov:1998ku,Ketov:2000zw}. 
For the moment the best-understood systems have one linear and one non-linear supersymmetry, 
since we have a clear understanding of the properties of the goldstone multiplet (which in the pure fermionic sector will always reduce to the Volkov--Akulov model \cite{Volkov:1973ix,Kuzenko:2011tj}). 
In particular it can be either an abelian N=1 vector multiplet with Born--Infeld dynamics \cite{Cecotti:1986gb,Bagger:1996wp,Kuzenko:2000tg,Kuzenko:2015rfx}, 
or a scalar N=1 multiplet (linear or chiral) \cite{Bagger:1997pi,Rocek:1997hi,Gonzalez-Rey:1998vtf}. 
These N=1 multiplets are embedded into N=2 superfields that satisfy some nilpotency constraints such that the second supersymmetry becomes non-linear (these constraints are of the same type as for the N=1 theories \cite{Rocek:1978nb,Lindstrom:1979kq,Casalbuoni:1988xh}). 
Beyond  the properties of these goldstino multiplets, 
some work has been done to understand the coupling of more than one abelian multiplet, 
possibly satisfying further constraints \cite{Ferrara:2014oka,Ferrara:2016crd}. 
For systems with fully broken extended supersymmetry one imposes further constraints, 
as studied e.g. in \cite{Kuzenko:2011ya,Cribiori:2016hdz,Dudas:2017sbi,Kuzenko:2017gsc}.

One open question within the regime of 4D N=2 with partial non-linear supersymmetry is what happens when non-abelian gauge fields are included in the system. 
Such a question, apart from being of interest in its own right, is also relevant for the description of stacks of D-branes; 
which in a supersymmetric configuration will give rise to a $U(N)$ theory living on the world-volume of the $N$ coinciding branes. 
It is well-known that the dynamics of slowly varying electric fields on D-branes is described by the Born--Infeld action \cite{Metsaev:1987qp,Tseytlin:1999dj} (see also \cite{Bergshoeff:2013pia}) 
and one can ponder on the non-abelian generalization, which is still elusive \cite{Tseytlin:1997csa,Ketov:2000fv,CiriloLombardo:2005yy}. 
Interestingly, 
when the 4D N=1 abelian theory is required to have a second non-linear supersymmetry with the gaugino as the goldstone field, 
then the induced non-linear structure of the bosonic sector is fully captured by the Born--Infeld action \cite{Bagger:1996wp}. 
One can then investigate what happens when the 4D N=1 non-abelian theory is required to have a second non-linear supersymmetry. 
Our work here addresses precisely this gap.

The rest of the article is organized as follows. 
In the second section we review the Bagger--Galperin construction, 
while in the third section we proceed to analyze how the partial breaking is mediated to non-abelian multiplets, 
and study the decoupling of an N=1 heavy {\it superpartner} multiplet. 
In the fourth section we implement the appropriate constraint and find that the second non-linear supersymmetry does not enforce a Born--Infeld structure in the non-abelian sector, which can remain quadratic in the non-abelian field strenghts in the bosonic sector. 
In the fifth section we conclude and discuss our findings. 
It is worth stressing that our results indicate that in contrast to the partial breaking in the abelian case, 
in the non-abelian case the Born--Infeld structure is not directly revealed just from the requirement of the one extra non-linear supersymmetry; we discuss some limitations in our analysis that point towards ways to overcome this barrier. 
The article includes an appendix where we explain our notation and conventions.

\section{The Bagger-Galperin construction}

Here we briefly review the properties of the Bagger--Galperin construction \cite{Bagger:1996wp} and the logic behind it. 
The same logic  will then guide us in the non-abelian sector. 
As a first step we present the full 4D N=2 multiplet that captures the partial breaking of supersymmetry, and then we impose and analyze the nilpotency constraint that makes the second supersymmetry non-linear.

With our conventions (see the appendix) the N=2 superfield that leads to the partial breaking is built from two N=1 chiral superfields $X$ and $\Lambda^\alpha$, 
where 
\bea
\Lambda^\alpha = i \overline D^2 D^\alpha U \,. 
\eea
Here $U$ is a 4D N=1 abelian vector superfield and the $D_\alpha$ are the superspace derivatives of the N=1 supersymmetry - the one that is actually preserved and manifest. 
The component fields are defined by superspace projection as usual 
\bea
D_\alpha \Lambda_\beta | = f_{\alpha \beta} - i C_{\alpha \beta} d \quad , \quad \Lambda_\alpha | =  \lambda_\alpha \,, 
\eea
where $f_{\alpha \beta}$ is the Maxwell tensor in spinor index notation (see appendix for details on conversion to vector notation), 
the $d$ is a real scalar field that serves as the auxiliary field in the standard two-derivative theory, 
and $\lambda_\alpha$ is the gaugino. 
The second supersymmetry relates the two N=1 superfields as follows 
\bea
\label{dXL}
\delta X = -i \Lambda^\alpha \epsilon_\alpha \quad , \quad  
\delta \Lambda^\alpha  =  i (m + \overline D^2 \overline X) \epsilon^\alpha 
+ \partial^{\alpha \dot \alpha} X \overline \epsilon_{\dot \alpha} \,. 
\eea
Here we follow \cite{Hull:1985pq}, and let us note for the reader's convenience that we use the supersymmetry algebra $\{D_\alpha,\overline D_{\dot \alpha}\} = i \partial_{\alpha \dot \alpha}$. 
This second supersymmetry closes off-shell and so can be used to construct any type of off-shell action (it can also be used to analyze nilpotency constraints). 
A Lagrangian that can describe the dynamics of the system is 
\bea\label{Lag1}
{\cal L} = \int d^4 \theta X \overline X + \left\{ \int d^2 \theta \left( \frac12 \Lambda^2 + f X \right) + cc \right\} \,, 
\eea
which is invariant under both supersymmetries up to boundary terms, 
and we have made use of the convention that $\int d^2 \theta = D^2|$ and that $\Lambda^2 = \frac12 \Lambda^\alpha \Lambda_\alpha$. 
Here $f$ is a real constant which will be fixed momentarily so that we get proper kinetic terms for the vector multiplet once the second supersymmetry becomes non-linear.

The non-linear realization of supersymmetry can be introduced systematically following closely the intuition that we have from 4D N=1 \cite{Rocek:1978nb,Lindstrom:1979kq,Casalbuoni:1988xh}. 
Indeed it is now well-understood that to turn a broken supersymmetry into a non-linear realization one imposes appropriate nilpotency constraints \cite{DallAgata:2016syy}. 
Similarly here, to turn the second supersymmetry non-linear we impose the nilpotency constraint 
\bea
\label{X2}
X^2 = 0 \,. 
\eea
This leads to the recursive constraint of Bagger--Galperin \cite{Bagger:1996wp} by acting twice with the broken second supersymmetry. 
Indeed, acting once gives $X \Lambda^\alpha = 0$ and then acting once more gives $\delta (X \Lambda^\alpha) = 0$. 
This last condition implies 
\bea
X = - \frac{\Lambda^2}{m + \overline D^2 \overline X} \,. 
\eea
The full solution is found following the method of \cite{Bagger:1996wp} and reads 
\bea
\label{BG-sol}
X = - \frac{\Lambda^2}{m} 
- \frac{2}{m} \overline D^2 \left[ \frac{\Lambda^2 \overline \Lambda^2}{m^2 - \alpha + \sqrt{m^4 -2 \alpha m^2 - \beta^2}} \right] , 
\eea
where $\alpha$ and $\beta$ are real expressions defined by 
\bea
2 D^2 \Lambda^2 = \alpha + i \beta \,. 
\eea
Once we implement the constraint the full Lagrangian reduces to 
\bea
\label{LX}
{\cal L}_X = \left( -\frac12 m + f \right) \int d^2 \theta X + cc  \,.  
\eea
The constants are chosen freely with the only restriction that the kinetic term of the vector should be canonical, 
that is $2f/m <1$, 
which allows the typical choice $f=0$ and $m>0$, 
which we will also assume from now on. 
Once we evaluate the bosonic sector and integrate out the auxiliary fields the Lagrangian \eqref{LX} delivers the Born--Infeld Lagrangian. 
In particular the auxiliary field of the vector multiplet, $d$, has the property 
\bea
\frac{\partial {\cal L}_X}{\partial d} \Big{|}_{d=0} = 0 \,, 
\eea
up to fermions. 
Replacing the on-shell values for the auxiliary field in \eqref{LX} one finds 
\bea
{\cal L}_X\Big{|}_B^{(aux. \ on-shell)} \equiv {\cal L}_{\rm Born-Infeld}. 
\eea

\section{Non-abelian N=2 and mediation of the partial breaking}

Now we turn to the non-abelian sector which we choose to be described by the k\"ahlerian vector multiplet (see \cite{Gates:1983py,deWit:1983xhu,Gunaydin:1983bi}). 
The supersymmetry transformations are taken directly from \cite{Hull:1985pq} and are off-shell. 
They act on N=1 superfields and they transform the chiral superfield to the vector superfield and back. 
We have 
\bea
\label{Phi-susy}
\delta \Phi &=& -i W^\alpha D_\alpha \varepsilon \,, 
\\
\label{V-susy}
e^{-V} \delta e^V &=& \overline \varepsilon \Phi + \varepsilon  e^{-V} \overline \Phi e^V \,, 
\eea
where $W^\alpha = i \overline D^2 \left( e^{-V} D^\alpha e^V  \right)$. 
Here these non-abelian fields all have an expansion in the adjoint representation of the form $\Phi = \Phi^A T^A$ and $V=V^A T^A$, 
where $A$ are Lie algebra indices, while the generators obey the normalization: ${\rm tr}  T^A T^B = \delta^{AB}$.\footnote{In general the trace of $T^A T^B$ will be proportional to the Killing metric which in turn will be proportional to delta with some coefficient.}  
The scalar superspace parameter $\varepsilon$ contains the supersymmetry transformation in the superspace expansion 
$\epsilon_\alpha = D_\alpha \varepsilon |$, 
while $\overline D_{\dot \alpha} \varepsilon = 0 = \partial_{\alpha \dot \alpha} \varepsilon$. 
The simplest Lagrangian invariant under the N=2 supersymmetry transformations is 
\bea
\label{Ltr}
{\cal L}_{NA} = \frac12  {\rm tr} \left[ 
\int d^2 \theta  W^2 + c.c. 
\right] 
+ {\rm tr} \int d^4 \theta \Phi e^{-V} \overline \Phi e^V \,, 
\eea
with the trace over the group indices. 
Note that the gauge transformations at the superspace level take the form 
\bea 
\label{GTS}
\Phi \to e^{i S} \Phi e^{-i S} 
\quad , \quad 
e^V \to e^{i \overline S} e^V e^{-i S} 
\quad , \quad 
W_\alpha \to e^{i S} W_\alpha e^{-i S} \,, 
\eea 
where $S = S^A T^A$ is the chiral superfield parametrizing the gauge transformations. 
From \eqref{GTS} one also deduces the transformations of the complex conjugate quantities.  
Perturbatively, the Lagrangian \eqref{Ltr} describes massless fields, that,  
in the bosonic sector correspond to gauge fields of the non-abelian N=1 gauge multiplet 
and scalar fields of the N=1 chiral multiplet transforming in the adjoint representation of the non-abelian group. 
Clearly, the only (perturbative) way for the gauge fields to get a mass is through a Brout--Englert--Higgs mechanism, 
whereas the scalars can get a gauge invariant mass if the N=2 is partially broken to N=1.

At this point we are ready to discuss the elimination of the 4D N=1 chiral superfield sector of the N=2 multiplet with the use of appropriate constraints. 
We will do this in two steps: 
First we deduce the appropriate constraint and then we solve it. 
To deduce/motivate the constraint that eliminates the scalar superfield non-abelian N=1 sector of the N=2 multiplet 
we rely on the intuition we have from 4D N=1 non-linear realizations and the analysis of the properties and origin of constraints given in \cite{DallAgata:2016syy}; 
that is, the term that leads to a large mass to some components should guide us to find the constraint that eliminates the given components. 
The combination of superspace terms that are invariant under the N=2 and introduce a mass for the non-abelian N=1 chiral multiplets, while the gauge sector remains massless, are 
\bea
{\cal L}_\gamma = \gamma \times {\rm tr} \left[ \int d^2 \theta \left( m\Phi^2 + 2 X W^2 + 2 \Phi W^\alpha \Lambda_\alpha \right) 
+ \int d^4 \theta \left( \Phi^2 \overline X + 2X \Phi e^{-V} \overline \Phi e^V \right) 
\right] + c.c. 
\eea
These terms essentially mediate the partial breaking to the non-abelian sector. 
Indeed the first term clearly describes the standard well-defined mass for a chiral multiplet which here would be of order $\gamma m$. 
Now imagine that we take the formal limit $\gamma \to \infty$, 
which, even though it may not be physically consistent per se, 
allows us to get intuition for how the low energy effective theory behaves when the mass of the N=1 chiral sector is large. 
Clearly when the mass is large the heavy chirals are integrated out and the point is to remove them in a supersymmetric way. 
One can convince oneself that the superspace equations of motion for $\Phi$ have the form 
\bea
2 m \Phi + 2 W^\alpha \Lambda_\alpha 
+ 2 \Phi \overline D^2 \overline X  
+ 2 \overline D^2 \left[ (X + \overline X) e^{-V} \overline \Phi e^V \right] 
+ \gamma^{-1} \times \left[\text{other terms}\right] = 0 \,, 
\eea
where we have divided by $\gamma$. 
Once we let $\gamma \to \infty$ we deduce a constraint for $\Phi$ that reads 
\bea
\label{DERV}
m \Phi + W^\alpha \Lambda_\alpha 
+ \Phi \overline D^2 \overline X  
+ \overline D^2 \left[ (X + \overline X) e^{-V} \overline \Phi e^V \right] = 0 \,. 
\eea
This constraint is actually the one that we will be eventually solving (albeit re-derived in a different way), 
but we can get a more intuitive form for its meaning by multiplying with $X$. 
Due to the nilpotency properties of $X$, after some manipulations this yields
\bea
\label{C1}
X \Phi = 0 \,. 
\eea
To deduce this condition one actually first multiples \eqref{DERV} with both $X$ and $\overline X$, 
to conclude that $X \overline X \Phi =0$ for $m\ne0$, and then the $\Phi X=0$ easily follows by multiplying \eqref{DERV} only with $X$.

These type of constraints of course have been studied for a 
multitude of N=2 abelian gauge theories in \cite{Ferrara:2014oka,Ferrara:2016crd}, 
and for standard 4D N=1 chiral superfields in \cite{Brignole:1997pe}. 
It is therefore known that they eliminate the lowest component of the given superfield, 
and that they can be applied self-consistently.

\section{Non-abelian theories with partially non-linear SUSY}

Since we have found the appropriate constraint \eqref{C1} that eliminates the superfields $\Phi^A$ we can now proceed to solve it. 
Here we are thinking of the constraint as an inherent property of the system not as a description of a low energy effective theory; 
it may be so, but we are not restricted to that. 
The upshot of this section is that we uncover the structure of the Super-Yang--Mills theory when it has a second non-linearly realized supersymmetry, and present a simple Lagrangian.

\subsection{Solving the constraint}

We start by solving the constraint \eqref{C1} in a supersymmetric way. 
By acting once with the broken supersymmetry transformations (namelly \eqref{dXL}, \eqref{Phi-susy} and \eqref{V-susy}) 
on \eqref{C1} we deduce 
\bea
X W^\alpha + \Phi \Lambda^\alpha = 0 \,. 
\eea
Acting once more we find 
\bea 
\label{Phi1}
\Phi = - \frac{W^\alpha \Lambda_\alpha}{m + \overline D^2 \overline X} 
- \frac{X}{m + \overline D^2 \overline X} \overline D^2 \left( e^{-V} \overline \Phi e^V \right) \,.  
\eea
Notice that \eqref{Phi1} is identical to \eqref{DERV} up to $\overline \Phi \overline X$ terms which in any case vanish due to \eqref{C1}, 
therefore we proceed with the form \eqref{Phi1}. 
The complex conjugate expression for \eqref{Phi1} reads 
\bea
\label{bPhi}
\overline \Phi = - \frac{\overline W^{\dot \alpha} \overline \Lambda_{\dot \alpha}}{m + D^2 X} 
- \frac{\overline X}{m + D^2 X} D^2 \left( e^{V} \Phi e^{-V}\right) \,.  
\eea 
To proceed towards the solution we first define the following two anti-chiral superfields (from the N=1 perspective) 
\bea
\label{RM}
R = D^2 \left( e^{V} \Phi e^{-V}\right) \quad , \quad M = m + D^2 X \,, 
\eea
which allow us to write the expressions in a more compact form. 
We can equally well define their chiral counterparts $\overline R$ and $\overline M$ 
by complex conjugation. 
Using the complex conjugate expression \eqref{bPhi} allows us to iterate once more and bring \eqref{Phi1} to the form 
\bea
\Phi = - \frac{W^\alpha \Lambda_\alpha}{\overline M} 
+ \frac{X}{\overline M} \overline D^2 \left( e^{-V}  \frac{\overline W^{\dot \alpha} \overline \Lambda_{\dot \alpha}}{M}  e^V \right)
+ \overline D^2 \left( e^{-V}  \frac{X \overline X}{M \overline M} R e^V \right) \,. 
\eea
From here we see that the final expression we actually need to evaluate to find the full expression for the composite $\Phi$ is simply given by $X \overline X R$. 
Clearly, since we can think of all the composite superfields as having a power-series expansion in terms of the $\Lambda$ superfields, also $R$ unavoidably has a $\Lambda$ expansion of the form 
\bea
\label{Reff}
R = R_{eff.} + {\cal O}(\Lambda) 
\eea
which means that in the recursive formula only the first piece enters, 
since $X \overline X \sim \Lambda^2 \overline \Lambda^2$. 
Therefore we only need to find an explicit expression for the part of $R$ that is independent of the $\Lambda$, 
that is the part that we call $R_{eff.}$. 
The same methodology was also followed in \cite{Bagger:1996wp} where it led to the square root structure in \eqref{BG-sol}. 

To deduce the recursive equation only in terms of $R_{eff.}$ we go through the following steps. 
First we turn \eqref{Phi1} into 
\bea 
\label{Phi2}
e^{V} \Phi e^{-V}  = - e^{V}  \frac{W^\alpha \Lambda_\alpha}{\overline M} e^{-V}
- e^{V}  \frac{X}{\overline M} \overline R e^{-V} \,, 
\eea
by inserting $R$ from \eqref{RM} and by multiplying the full equation from the left and from the right with appropriate $e^V$ factors. 
Then we act with $X \overline X D^2$ on both sides which gives 
\bea
\label{XXbD2}
|X|^2 D^2 \left( e^{V} \Phi e^{-V} \right) 
= |X|^2 D^2 \left( - e^{V}  \frac{W^\alpha \Lambda_\alpha}{\overline M} e^{-V} \right) 
- |X|^2 D^2 \left( e^{V}  \frac{X}{\overline M} \overline R e^{-V} \right) . 
\eea
Due to the nilpotency condition \eqref{X2} the superspace derivatives are forced to act on specific superfields, 
or combinations of superfields, otherwise the full contribution vanishes. 
Therefore \eqref{XXbD2} reduces to  
\bea 
\label{XXbR}
|X|^2 R
= |X|^2 D^2 \left( - e^{V}  \frac{W^\alpha \Lambda_\alpha}{\overline M} e^{-V} \right) 
- |X|^2 \frac{D^2 X}{\overline M} e^{V}   \overline R e^{-V} \, , 
\eea
where we have once more inserted $R$ in the appropriate places. 
From here we derive the equation for the effective part of $R$, 
which, using the formula \eqref{Reff}, takes the form 
\bea 
\label{EqReff}
R_{eff.}
= D^2 \left( - e^{V}  \frac{W^\alpha \Lambda_\alpha}{\overline M} e^{-V} \right) 
- \frac{D^2 X}{\overline M} e^{V}   \overline R_{eff.} e^{-V} \, . 
\eea

Let us pause here and justify the step from \eqref{XXbR} to \eqref{EqReff}. 
First we notice that all the terms in \eqref{XXbR} are multiplied with $X \overline X \sim \Lambda^2 \overline \Lambda^2$. 
This is crucial, 
because one can see that the following equivalence generically holds (see e.g. \cite{DallAgata:2016syy}) 
\bea 
\label{LBLC}
\Lambda^2 \overline \Lambda^2 B = \Lambda^2 \overline \Lambda^2 C 
\quad \Leftrightarrow \quad B = C + {\cal O}(\Lambda), 
\eea
where here $B$ and $C$ are some N=1 superfields, 
which of course should also have an assigned transformation under the broken supersymmetry. 
Notice that the equivalence \eqref{LBLC} still holds perfectly well even if either $B$ or $C$ 
or both have some extra dependence on $\Lambda$. 
This property can also be understood from the properties of non-linear supersymmetry since $\Lambda$ is nothing but the goldstino superfield for the broken supersymmetry. 
Therefore, taking into account \eqref{LBLC}, equation \eqref{XXbR} reduces to 
\bea 
\label{RRL}
R = D^2 \left( - e^{V}  \frac{W^\alpha \Lambda_\alpha}{\overline M} e^{-V} \right) 
- \frac{D^2 X}{\overline M} e^{V}   \overline R e^{-V} + {\cal O}(\Lambda) \, . 
\eea
However, due to \eqref{Reff}, 
the $R$ in equation \eqref{RRL} might as well be directly replaced with $R_{eff.}$, 
since the difference is always taken care of by the ${\cal O}(\Lambda)$ terms. 
Finally, 
since for the $R_{eff.}$ we do not need to explicitly keep track of any bare ${\cal O}(\Lambda)$ terms we can simply drop the ${\cal O}(\Lambda)$ altogether and therefore arrive at equation \eqref{EqReff}. 
Notice that if one pushes the $D^2$ in equation \eqref{EqReff} to act in the various terms in the parentheses then it will give bare ${\cal O}(\Lambda)$ terms; 
these terms are not relevant for us and one can also drop them, 
but it is more convenient to keep the formula in the compact form where the $D^2$ stays outside. 
Indeed, once we multiply with $X \overline X$ these terms will automatically drop out. 
In other words one can consider that $R_{eff.}$ is always defined up to ${\cal O}(\Lambda)$ terms, 
therefore equation \eqref{EqReff} is perfectly fine as it stands.

Now we proceed to determine $R_{eff}.$ from equation \eqref{EqReff}, 
which clearly has to be solved recursively. 
From the right-hand-side of \eqref{EqReff} we see that we need to use its complex conjugate, 
which has the form 
\bea
\overline R_{eff.}
= \overline D^2 \left( - e^{-V}  \frac{\overline W^{\dot \alpha} \overline \Lambda_{\dot \alpha}}{M} e^{V} \right) 
- \frac{\overline D^2 \overline X}{M} e^{-V} R_{eff.} e^{V} \, . 
\eea
Then combining \eqref{EqReff} with its complex conjugate gives 
\bea \nonumber
\label{Reff-fin}
R_{eff.}
&=& D^2 \left( - e^{V}  \frac{W^\alpha \Lambda_\alpha}{\overline M} e^{-V} \right) 
\\\nonumber
&& + \frac{D^2 X}{\overline M} e^{V} 
\overline D^2 \left( e^{-V}  \frac{\overline W^{\dot \alpha} \overline \Lambda_{\dot \alpha}}{M} e^{V} \right)  e^{-V} 
\\  
&& + \frac{D^2 X}{\overline M} e^{V}   \frac{\overline D^2 \overline X}{M} e^{-V} R_{eff.} e^{V} e^{-V} \,, 
\eea
which signals the end of the recursive calculation because we have a bare $R_{eff.}$ in both sides, 
albeit multiplied with a variety of coefficients. 
Solving then \eqref{Reff-fin} algebraically gives our final expression for the $R_{eff.}$, 
which reads 
\bea
\label{Reff-res}
R_{eff.} = \frac{ 
D^2 \left( - e^{V}  \frac{W^\alpha \Lambda_\alpha}{\overline M} e^{-V} \right) 
+ \frac{D^2 X}{\overline M} e^{V} 
\overline D^2 \left( e^{-V}  \frac{\overline W^{\dot \alpha} \overline \Lambda_{\dot \alpha}}{M} e^{V} \right)  e^{-V}
}{ 1 - \frac{D^2 X}{\overline M} \frac{\overline D^2 \overline X}{M} } \,. 
\eea
Therefore our final result for $\Phi$ is given by 
\bea
\label{Phi-fin}
\Phi = - \frac{W^\alpha \Lambda_\alpha}{\overline M} 
+ \frac{X}{\overline M} \overline D^2 \left( e^{-V}  \frac{\overline W^{\dot \alpha} \overline \Lambda_{\dot \alpha}}{M}  e^V \right)
+ \overline D^2 \left( e^{-V}  \frac{X \overline X}{M \overline M} R_{eff.} e^V \right) \,, 
\eea
with $R_{eff.}$ given by \eqref{Reff-res}. 
We see that the full expression for $\Phi$ has a formidable structure. 
This happens because it keeps track of the underlying second non-linear supersymmetry which now acts on the $W_\alpha$ superfield transforming it into itself (and goldstini) via equation \eqref{V-susy}. 
The pure bosonic sector of $\Phi$ is simpler, especially when we incorporate it into a Lagrangian and integrate out the auxiliary fields. 
We will do this right away.

\subsection{The bosonic Lagrangian sector}

We will work with the non-abelian Lagrangian that has the superspace form given in \eqref{Ltr}. 
Of course a full model should contain both the non-abelian sector ${\cal L}_{NA}$ given by \eqref{Ltr} together with the goldstino Lagrangian ${\cal L}_X$ given in \eqref{LX}, 
that is we consider 
\bea
\label{Lfull}
{\cal L} = {\cal L}_X + {\cal L}_{NA} \,. 
\eea
A consistent Lagrangian needs both terms otherwise the goldstino sector does not have a standard kinetic term. 
The full component form of the Lagrangian \eqref{Lfull} will lead to an unwieldy expression due to the highly non-linear fermionic sector, therefore we will only target the purely bosonic contributions. 
Specifically, the bosonic sector of the full Lagrangian \eqref{Lfull} is included in the terms 
\bea 
\label{Bos1}
{\cal L}_B = \frac12  {\rm tr}  \left[ D^2  W^2  + \overline D^2  \overline W^2 \right] \Big{|}_B 
+  {\rm tr} \, D^2 \Phi \overline D^2 \overline \Phi \Big{|}_B + {\cal L}_X \Big{|}_B \,. 
\eea
The first term in \eqref{Bos1} is the standard kinetic term for the super Yang--Mills sector, 
the second term, due to the composite nature of $\Phi$, 
will give rise to a series of non-linear terms that will include also the super Yang--Mills sector, 
and the last term is essentially nothing but the Born--Infeld $U(1)$ sector (once the auxiliary fields are integrated out of course).

To turn to components we use the conventions for the definition of the bosonic component fields which yield in particular 
\bea
\label{contract} 
D_\alpha W_\beta | = {\cal F}_{\alpha \beta} - i C_{\alpha \beta} {\cal D} \quad , \quad  
D^2 (W^\alpha \Lambda_\alpha)|_B =- {\cal F}^{\alpha \beta} f_{\alpha \beta} + 2 {\cal D} d \, , 
\eea
where ${\cal F}_{\alpha \beta} = {\cal F}_{\alpha \beta}^A T^A$ corresponds to the Yang--Mills Maxwell field strength in spinor notation and ${\cal D} = {\cal D}^A T^A$ corresponds to the auxiliary fields, which are also in the adjoint representation. 
Then the bosonic sector of the first term in \eqref{Bos1} in particular is 
\bea 
\nn
\frac12 {\rm tr}  \left[ D^2  W^2  + \overline D^2  \overline W^2 \right] \Big{|}_B 
&=& -\frac14  {\rm tr}  \left[ {\cal F}^{\alpha \beta} {\cal F}_{\alpha \beta} 
+ {\cal F}^{\dot \alpha \dot \beta} {\cal F}_{\dot \alpha \dot \beta}  \right] 
+ {\rm tr}  {\cal D} {\cal D}  
\\
\label{Bos2}
&=& -\frac14  \left[ {\cal F}^{\alpha \beta}{}^A {\cal F}_{\alpha \beta}^A 
+ {\cal F}^{\dot \alpha \dot \beta}{}^A {\cal F}_{\dot \alpha \dot \beta}^A  \right] 
+  {\cal D}^A {\cal D}^A \,. 
\eea 
To go from spinor index notation to vector notation for the Lorentz indices we refer the reader to the appendix. 
To evaluate the second term of \eqref{Bos1} we only need to derive 
\bea
\label{D2PhiB}
D^2 \Phi |_B = \frac{M D^2 (W^\alpha \Lambda_\alpha) + D^2 X \overline D^2 (\overline W^{\dot \alpha} \overline \Lambda_{\dot \alpha})}{M \overline M - D^2 X \overline D^2 \overline X}  \Big{|}_B \,, 
\eea
where the terms $D^2 (W^\alpha \Lambda_\alpha)$ are given in \eqref{contract}.

To reduce the full bosonic sector to the physical/propagating contributions we should integrate out all the auxiliary fields. 
Since at this stage we are only interested in the bosonic sector we will ignore all fermionic contributions to the on-shell value of the auxiliary fields. 
From an inspection of the component form of the Lagrangian \eqref{Bos1} one can see that the auxiliary fields always appear at least quadratically, 
and so their equations of motion are bound to have at least one solution that takes the form
\bea 
\label{dD0}
d = 0 = {\cal D} \,, 
\eea
always up to fermions. 
Since the solution \eqref{dD0} has vanishing VEVs for the auxiliary fields it means that it preserves supersymmetry 
and therefore it is in any case stable, even if other solutions to the auxiliary field equations exist. 
Eventually the term that produces the higher-order interactions within \eqref{Bos1} takes the form  
\bea 
  {\rm tr} \, D^2 \Phi \overline D^2 \overline \Phi \Big{|}_B^{(aux. \ on-shell)} = 
  \frac{ |m + 2 D^2 X|^2 \, {\rm tr} \, K^2 + m^2 \, {\rm tr} \, L^2 - 4 i m B \, {\rm tr} \, KL }{(m^2 + 2 m A )^2} \,, 
\eea
where we have defined 
\bea
{\cal F}^{\alpha \beta} f_{\alpha \beta}  = K + i L \quad , \quad  
A = \frac12 (D^2 X + \overline D^2 \overline X )| \quad , \quad 
B =  \frac12 (D^2 X - \overline D^2 \overline X )| \, . 
\eea
With these formulas we have the ingredients of the component field bosonic sector, 
but we will not analyze it in detail. 
However, to get a feeling for the form of the higher order interactions, 
we expand the composite terms. 
For example the mixing between the non-abelian and the abelian sector up to forth order has the form 
\bea
\frac{1}{m^2} \left( {\rm Tr} K^2 + {\rm Tr} L^2 \right) 
\sim \frac{1}{m^2} {\cal F}^A_{mn} {\cal F}^A_{kl} f^{mn} f^{kl} 
+ \frac{1}{m^2} {\cal F}^A_{mn} {\cal F}^A_{kl} f^{*mn} f^{*kl}  \,. 
\eea

Taking now into account our results we see that central to the final on-shell form of the full Lagrangian \eqref{Lfull} is that its bosonic sector does not contain any Yang--Mills field strengths beyond the second order. 
This means that, in contrast to the pure abelian case, 
simply requiring to have non-abelian theories with partially broken and non-linearly realized supersymmetry does not buy us the Born--Infeld structure. 
In the Discussion section we evaluate what might be the underlying reason hindering the appearance of the non-abelian Born--Infeld structure from our analysis.

Let us note for completeness that if instead of the non-abelian sector we had an {\it abelian} extra sector 
(described again by 4D N=1 superfields $\Phi$ and $W_\alpha$), 
the nilpotency constraint \eqref{C1} could still be applied, 
and the solution \eqref{Phi-fin} would essentially have the same form. 
Then our final results for the Lagrangian would still have the same form as the results of this section 
- without traces, of course.

\section{Discussion}

In this work we studied how the partial non-linear realization of supersymmetry works for 4D N=2 Super Yang--Mills theories described by the k\"ahlerian vector multiplet. 
The {\it goldstino} sector is described by the Super Born--Infeld sector, 
and with appropriate supersymmetry-preserving constraints we have mediated the non-linear supersymmetry to the non-abelian sector. 
Our work can be summarized as two main achievments: 
First, we were able to solve the appropriate constraints and express them in closed form paving the way for future studies. 
Second, we have seen that the non-abelian extension of the Born-Infeld action did not emerge just by the requirement of having a second non-linear supersymmetry in the non-abelian sector.

Our work has laid the foundation for a variety of future works and has raised various questions that deserve further study - 
we will highlight a few: \\
Firstly, In this work all the N=1 chiral multiplets which have scalars in the adjoint are set to have vanishing VEVs due to the constraint $X \Phi=0$; 
An interesting development is to write down the proper theory with partial non-linear supersymmetry where the VEVs of the lowest components of the scalars fields will be non-trivial. \\
Secondly, since in our final Lagrangians we found that all non-abelian field-strengths in the bosonic sector stop at second order, it is an interesting technical question on its own right to construct higher order interactions for the non-abelian sector that still preserve half linear and half non-linear supersymmetry. \\ 
Thirdly, 
one can also turn to {\it hypermultiplets}, which can in principle accommodate both the supersymmetry breaking sector with appropriate constraints \cite{Bagger:1997pi,Rocek:1997hi} but are also the appropriate object to describe matter fields that transform in the fundamental or in some other representation. 
Such an option for the matter sector would allow to go beyond the adjoint representation, to which by construction our fields belonged here.

Our analysis and our discussion until here directly raises the question of how the higher order terms naturally appear in a {\it non-abelian} Born--Infeld form while preserving partially non-linear supersymmetry. 
One can speculate about the possible ways to proceed. 
One ingredient that may be central in such pursuit, which was missing in our analysis here, 
is the fact that we do not have a distinct source for supersymmetry breaking within the non-abelian sector, 
say $\tilde m$. 
We believe this is one of the most interesting directions to explore. 
For simplicity one can start from an example with two abelian fields with two different breaking sectors to avoid the extra intricacy that the non-abelian structure will bring.

\section*{Acknowledgements}

We would like to thank Martin Ro\v{c}ek and Augusto Sagnotti for feedback and comments. 
U.L. thanks the Department of Physics of the National Technical University of Athens for hospitality during the preparation of this work.

\appendix 

\section{Notation and conventions}

We use the 4D superspace conventions of \cite{Book}. Here is a brief summary of some relevant features. 

Our Yang-Mills indices are upper case while lower case latin indices are Lorentz indices. 
We use two-component Weyl spinor indices which are raised and lowered using the second Pauli matrix, $C_{\al\be}=(\sigma^2)_{\al\be}$,  
making use of the convention 
\begin{equation}
\psi^\al=C^{\al\be}\psi_\be~, \qquad \psi_\al=\psi^\be C_{\be\al} \, , 
\end{equation}
and idem for dotted indices. 
Important properties of the spinor metric $C$ are:  
\bea\nonumber
&&C_{\alpha\beta}C^{\gamma\delta}=\delta^\gamma_\alpha\delta^\delta_\beta-\delta^\gamma_\alpha\delta^\beta_\alpha\quad\Rightarrow ~~C^{\gamma\delta}C_{\gamma\beta}=\delta^\delta_\beta~~\makebox{and}~~ C^{\gamma\delta}C_{\gamma\delta}=2 \,, 
\eea
while the standard identity for anti-commuting spinor indices reads: 
$\eta_\al\xi_\beta-\eta_\beta\xi_\al=C_{\beta\al}\eta^\gamma\xi_\gamma$. 
Using the Pauli $\sigma$ matrices, vector indices are represented by pairs of spinor indices, one undotted and one dotted:
\bea
V^m \to V^{\mu\dot \mu} :~ V^m=\frac 1 {\sqrt{2}}\sigma^m_{~~\mu\dot \mu}V^{\mu\dot \mu} \,, 
\eea
etc. 
For an antisymmetric field strength $F_{mn}$, we have
\bea\label{FS}
F_{mn} \to F_{\mu\dot\mu\nu\dot\nu}=C_{\nu\mu}\bar F_{\dot\mu\dot \nu}+C_{\dot\nu\dot\mu} F_{\mu\nu} \,, 
\eea
where the symmetric two spinor 
\bea
F_{\mu\nu}=\frac 12 F_{\mu~\nu\dot\nu}^{~\dot\nu} 
\eea
is the selfdual part and its complex conjugate is the anti-selfdual part of the field strength.
The four-dimensional Levi-Civita symbol takes the following spinorial form
\bea\label{LC}
\epsilon_{abcd} \to i\big( C_{\al\delta}C_{\beta\gamma}C_{\dot\al\dot\beta}C_{\dot\gamma\dot\delta}
-C_{\al\beta}C_{\gamma\delta}C_{\dot\al\dot\delta}C_{\dot\beta\dot\gamma}\big) \,. 
\eea
From \re{FS} we find
\bea 
F_{mn}F^{mn} \to 2\big(F_{\mu\nu}F^{\mu\nu}+\bar F_{\dot\mu\dot\nu}\bar F^{\dot\mu\dot\nu}\big) 
\eea
and from \re{LC} that
\bea
F_{mn}F^{*mn}=\frac 12F_{mn}\epsilon^{mnrs}F_{rs} \to -2i\big(F_{\mu\nu}F^{\mu\nu}-\bar F_{\dot\mu\dot\nu}\bar F^{\dot\mu\dot\nu}\big)\,. 
\eea
For further notational detail the reader is invited to consult sec.~\!$3.1$ of \cite{Book}.

\end{document}